\documentclass[twocolumn,showpacs,preprintnumbers]{revtex4-1}
\usepackage{graphicx}
\def \be{\begin{equation}}
\def \ee{\end{equation}}
\def \bdm{\begin{eqnarray}}
\def \edm{\end{eqnarray}}
\def \kappafl{\kappa_{\scriptscriptstyle F\!L}}
\begin{document}
\setlength{\abovedisplayskip}{6pt}
\setlength{\belowdisplayskip}{6pt}
\preprint{Submitted to Physics of Plasmas}
\title{Heuristic Description of Perpendicular Diffusion of Energetic Particles in Astrophysical Plasmas}
\author{A. Shalchi}
\email{andreasm4@yahoo.com}
\affiliation{Department of Physics and Astronomy, University of Manitoba, Winnipeg, Manitoba R3T 2N2, Canada}
\date{\today}
\begin{abstract}
A heuristic approach for collisionless perpendicular diffusion of energetic particles is presented. Analytic forms for the corresponding diffusion coefficient
are derived. The heuristic approach presented here explains the parameter $a^2$ used in previous theories in order to achieve agreement with 
simulations and its relation to collisionless Rechester \& Rosenbluth diffusion. The obtained results are highly relevant for applications because previously
used formulas are altered significantly in certain situations.
\end{abstract}
\maketitle
\section{Introduction}
A problem of utmost importance is the interaction between electrically charged particles and magnetized plasmas. It plays a significant
role in a variety of physical systems ranging from fusion devices, over the solar wind, to the shock fronts of supernova explosions.
In all of those scenarios energetic particles experience scattering due to complicated interactions with turbulent magnetic fields. Some early
work was done based on perturbation theory also known as quasi-linear theory (see, e.g., Jokipii (1966)) but in general this approach
fails. Some more heuristic arguments but also systematic theories have been developed focusing on electron heat transport in fusion
plasmas where collisions are assumed to play a significant role (see, e.g., Rechester \& Rosenbluth (1978), Kadomtsev \& Pogutse (1979),
and Krommes et al. (1983)). In space plasmas such as the solar wind or the interstellar medium, on the other hand, collisions are absent
and, thus, it was concluded that the aforementioned approaches are not applicable. The assumption of exponential field line separation
was also questioned (see, e.g., Matthaeus et al. (2003)). In the context of astrophysical plasmas, however, one still finds perpendicular
diffusion in most cases as shown via test-particle simulations (see, e.g., Giacalone \& Jokipii (1999) and Qin et al. (2002)) but it
remained unclear what the mechanisms behind this type of transport are. Progress has been achieved due to the development of the
non-linear guiding center theory (Matthaeus et al. (2003)), the unified non-linear transport (UNLT) theory of Shalchi (2010), as well
as its time-dependent generalization (Shalchi (2017)). Within \textit{diffusive UNLT theory} the perpendicular diffusion coefficient
is given by
\be
\kappa_{\perp} = \frac{a^2 v^2}{3 B_0^2} \int d^3 k \; \frac{P_{xx} (\vec{k})}{F(k_{\parallel},k_{\perp}) + (4/3) \kappa_{\perp} k_{\perp}^2 + v/\lambda_{\parallel}}
\label{UNLT}
\ee
where $F (k_{\parallel},k_{\perp}) = (v k_{\parallel})^2 /(3 \kappa_{\perp} k_{\perp}^2)$. The solution of this equation depends on the 
spectral tensor $P_{nm}$ describing the magnetic fluctuations, the parallel mean free path $\lambda_{\parallel} = 3 \kappa_{\parallel} / v$,
the particle speed $v$, and the mean field $B_0$. Asymptotic solutions of Eq. (\ref{UNLT}) and the importance of the Kubo number (Kubo (1963))
defined via $K = (\ell_{\parallel} \delta B_x) / (\ell_{\perp} B_0)$, depending on the parallel and perpendicular bendover scales $\ell_{\parallel}$
and $\ell_{\perp}$ as well as the turbulent magnetic field $\delta B_x$, have been discussed in Shalchi (2015). Eq. (\ref{UNLT}) shows good
agreement with most test-particle simulations in particular with those performed for three-dimensional turbulence with small and intermediate
Kubo numbers. Furthermore, Eq. (\ref{UNLT}) contains quasi-linear theory as well as the non-linear theory of field line random walk (FLRW)
developed by Matthaeus et al. (1995). In Shalchi (2017) \textit{time-dependent UNLT theory} has been derived which is represented by
\be
\frac{d^2}{d t^2} \langle (\Delta x)^2 \rangle = \frac{2 a^2}{B_0^2} \int d^3 k \; P_{xx} \big( \vec{k}, t \big)
\xi \big( k_{\parallel}, t \big) e^{- \frac{1}{2} \langle (\Delta x)^2 \rangle k_{\perp}^2}
\label{Generalsigmap}
\ee
with the parallel correlation function given by $\xi (k_{\parallel},t)=(v^2 / 3) (\omega_+ e^{\omega_+ t} - \omega_- e^{\omega_- t})/(\omega_+ - \omega_-)$
where $\omega_{\pm} = - v/(2 \lambda_{\parallel}) \pm [v^2/(2 \lambda_{\parallel})^2 - v^2 k_{\parallel}^2 / 3]^{1/2}$.
Eq. (\ref{UNLT}) can be derived from Eq. (\ref{Generalsigmap}) by employing a diffusion approximation. Furthermore, the theory explains why diffusion is restored and
this is entirely due to transverse complexity becoming important. Due to the exponential factor
in Eq. (\ref{Generalsigmap}), this means that diffusion is obtained if $\langle (\Delta x)^2 \rangle \geq 2 \ell_{\perp}^2$. However, there are
at least two remaining problems in the theory of perpendicular diffusion. First, there is a discrepancy between theory and simulations in the
large Kubo number regime which was previously balanced out by using the factor $a^2$ (see Eqs. (\ref{UNLT}) and (\ref{Generalsigmap}))
and by setting $a^2 = 1/3$ (see Matthaeus et al. (2003)). Furthermore, the question remains what the physics behind collisionless perpendicular diffusion is.
This letter provides an answer to both questions.

\section{The Three Rules of Perpendicular Diffusion}
We now formulate rules allowing us to derive formulas for the perpendicular diffusion coefficient without employing systematic theories.
Those rules are:
\begin{enumerate}
\item Perpendicular transport is only controlled by three effects, namely parallel transport, the random walk of magnetic field lines,
as well as transverse complexity. The last of these three effects leads to the particles getting scattered away from the original
magnetic field lines they were tied to.
\item We assume that the bendover scales $\ell_{\parallel}$ and $\ell_{\perp}$, the integral scales $L_{\parallel}$ and $L_{\perp}$,
the ultra-scale $L_U$, as well as the Kolmogorov scale $L_K$ are finite and non-zero. Furthermore, the parallel motion is assumed to
be ballistic at early times and thereafter turns into a diffusive motion described by the parallel diffusion coefficient $\kappa_{\parallel}$.
The FLRW is initially ballistic and becomes diffusive for larger distances. In this case it is described by the field line diffusion
coefficient $\kappafl$ which depends on some of the aforementioned scales.
\item 
In order to obtain normal diffusion, the particles need to leave the original magnetic field lines they followed. This happens
as soon as transverse complexity becomes significant corresponding to
\be
\big< \big( \Delta x \big)^2 \big> \geq 2 \ell_{\perp}^2.
\label{thecond}
\ee
It is assumed here that $\ell_{\perp}$ is the scale at which transverse complexity becomes significant. In principle this could be a different
scale such as the integral scale $L_{\perp}$. To use the bendover scale, however, is motivated by time-dependent UNLT theory (see Sect. 1).
What the perpendicular diffusion coefficient is depends solely on the state of parallel and field line transport at the time particles start
to satisfy condition (\ref{thecond}).
\end{enumerate}

\section{The Perpendicular Diffusion Coefficient}
In the following we construct the perpendicular diffusion coefficient $\kappa_{\perp}$ based on the three rules formulated above. We shall
derive eight cases which are summarized in Table \ref{TheCases}. As demonstrated, there are four different routes to perpendicular diffusion
as listed in Table \ref{TheRoutes}.

\begin{table*}
\caption{The eight cases of perpendicular transport. The results are compared with limits contained in time-dependent UNLT theory represented
by Eq. (\ref{Generalsigmap}).}
\renewcommand{\arraystretch}{1.8}
\begin{tabular}{|l|l|l|l|l|l|l|}
\hline
Case	& Parallel Motion	& Field Lines	& $\langle (\Delta x)^2 \rangle \geq 2 \ell_{\perp}^2$ 	& Perpendicular Transport & Diffusion Coefficient			& Described by UNLT Theory	\\
\hline
$1$ & Ballistic			& Ballistic		& No				& Ballistic 				& $d_{\perp} (t) = \frac{v^2}{3} \frac{\delta B_x^2}{B_0^2} t$				& Yes 						\\
$2$ & Ballistic			& Ballistic		& Yes				& Double-ballistic diffusion& $\kappa_{\perp}=\sqrt{\frac{2}{3}} v \ell_{\perp}\frac{\delta B_x}{B_0}$ 	& Yes 						\\
$3$ & Ballistic			& Diffusive		& No				& FLRW Limit 				& $\kappa_{\perp}=\frac{v}{2} \kappafl$										& Yes 						\\
$4$ & Ballistic			& Diffusive		& Yes				& FLRW Limit 				& $\kappa_{\perp}=\frac{v}{2} \kappafl$										& Yes 						\\
$5$ & Diffusive			& Ballistic		& No				& Fluid Limit				& $\kappa_{\perp}=\frac{\delta B_x^2}{B_0^2} \kappa_{\parallel}$	 		& Yes 						\\
$6$ & Diffusive			& Ballistic		& Yes				& Fluid Limit				& $\kappa_{\perp}=\frac{\delta B_x^2}{B_0^2} \kappa_{\parallel}$			& Yes 						\\
$7$ & Diffusive			& Diffusive		& No				& Compound Sub-diffusion 	& $d_{\perp} (t) = \kappafl \sqrt{\frac{\kappa_{\parallel}}{2 t}}$			& Only for small Kubo numbers	\\
$8$ & Diffusive			& Diffusive		& Yes				& CLRR Limit & $\kappa_{\perp}=\left( \frac{\kappafl}{\ell_{\perp}} \right)^2 \kappa_{\parallel}$		& Only for small Kubo numbers 	\\
\hline
\end{tabular}
\label{TheCases}
\end{table*}
\begin{table}
\caption{The four routes to perpendicular diffusion. In the first three cases perpendicular transport starts as ballistic motion which
then turns into a diffusive motion. In the fourth case the ballistic motion is followed by a sub-diffusive regime and 
thereafter diffusion is restored.}
\renewcommand{\arraystretch}{1.8}
\begin{tabular}{|l|l|l|}
\hline
Route									& Final State 					& Diffusion Coefficient																	\\
\hline
$1 \rightarrow 2$						& Double-ballistic diffusion	& $\kappa_{\perp}=\sqrt{\frac{2}{3}} v \ell_{\perp}\frac{\delta B_x}{B_0} $				\\
$1 \rightarrow 3 \rightarrow 4$			& FLRW Limit					& $\kappa_{\perp}=\frac{v}{2} \kappafl$													\\
$1 \rightarrow 5 \rightarrow 6$			& Fluid Limit					& $\kappa_{\perp}=\frac{\delta B_x^2}{B_0^2} \kappa_{\parallel}$						\\
$1 \rightarrow 7 \rightarrow 8$			& CLRR Limit 					& $\kappa_{\perp}=\left( \frac{\kappafl}{\ell_{\perp}} \right)^2 \kappa_{\parallel}$	\\
\hline
\end{tabular}
\label{TheRoutes}
\end{table}
\subsection{The Field Line Random Walk Limit}
First we assume that the random walk of magnetic field lines is diffusive in the scenario of interest
\be
\big< \big( \Delta x \big)^2 \big> = 2 \kappafl \big| z \big|.
\label{fldiff}
\ee
Note that the diffusion coefficient $\kappafl$ has length units.
If we assume that there are no collisions and no pitch-angle scattering, we can set $z = v \mu t$ where we used the pitch-angle cosine $\mu$.
Combining this with Eq. (\ref{fldiff}) and averaging over $\mu$ yields $\langle (\Delta x)^2 \rangle = v \kappafl t$ and, thus,
\be
\kappa_{\perp} = \frac{v}{2} \kappafl
\label{theflrwlimit}
\ee
corresponding to the FLRW limit. This limit is stable because if condition (\ref{thecond}) is met, it does not alter the transport. This case
is highly relevant in the limit of long parallel mean free paths corresponding to high particle energies (see Figs. \ref{GSexample} and \ref{NRMHDexample}).

\subsection{Compound Sub-diffusion}
If there is strong pitch-angle scattering the parallel motion is diffusive meaning that
\be
\big< \big( \Delta z \big)^2 \big> = 2 \kappa_{\parallel} t.
\label{paradiff}
\ee
Assuming that field lines are diffusive and particles follow field lines, we can combine Eqs. (\ref{fldiff}) and (\ref{paradiff}) to find
\be
\big< \big( \Delta x \big)^2 \big> \approx 2 \kappafl \sqrt{2 \kappa_{\parallel} t}.
\label{heuristiccompound}
\ee
The running perpendicular diffusion coefficient is then
\be
d_{\perp} (t) = \frac{1}{2} \frac{d}{d t} \big< \big( \Delta x \big)^2 \big>
\approx \kappafl \sqrt{\frac{\kappa_{\parallel}}{2 t}}
\label{running}
\ee
corresponding to sub-diffusive transport. However, diffusion will be restored as soon as condition (\ref{thecond}) is satisfied as discussed in the
next paragraph. For slab turbulence, on the other hand, this condition is never satisfied due to $\ell_{\perp} = \infty$ and thus, we find compound
sub-diffusion as the final state of perpendicular transport.

\subsection{The Collisionless Rechester \& Rosenbluth Regime}
We now assume that diffusion is restored as soon as the particles scatter away from their original field lines. This happens as soon as condition
Eq. (\ref{thecond}) is satisfied. We also assume that this happens after the particles travel the distance $L_K$ in the parallel direction leading to
$\kappa_{\perp}/\kappa_{\parallel} = \langle (\Delta x)^2 \rangle / \langle (\Delta z)^2 \rangle = \ell_{\perp}^2 / L_K^2$.
In order to eliminate $L_K$ we use the field line diffusion coefficient $\kappafl = \langle (\Delta x)^2 \rangle /(2 |z|) = \ell_{\perp}^2 / L_K$ yielding
\be
\kappa_{\perp} \approx \left( \frac{\kappafl}{\ell_{\perp}} \right)^2 \kappa_{\parallel}.
\label{theimportantformula}
\ee
Alternatively, one can replace the scale $\ell_{\perp}$ therein so that
\be
\kappa_{\perp} \approx \frac{\kappafl \kappa_{\parallel}}{L_K}
\label{useLK}
\ee
in agreement with equation (8) of Rechester \& Rosenbluth (1978) as well as equation (4) of Krommes et al. (1983). The quantity $L_K$ is either
called the \textit{Kolmogorov-Lyapunov length} or just the \textit{Kolmogorov length} (see, e.g., Krommes et al. (1983)). However, here $L_K$ is
not an \textit{exponentiation length} but a characteristic distance along the mean field at which transverse complexity becomes significant.
Furthermore, Eqs. (\ref{theimportantformula}) and (\ref{useLK}) were obtained without assuming collisions and, thus, we call this result the
\textit{collisionless Rechester \& Rosenbluth (CLRR) limit}. One can also obtain this by using a slightly different derivation. We
assume that we find compound sub-diffusion until the particles satisfy condition (\ref{thecond}) which happens at the diffusion time $t_d$
so that Eqs. (\ref{heuristiccompound}) and (\ref{running}) become $2 \ell_{\perp}^2 = 2 \kappafl \sqrt{2 \kappa_{\parallel} t_d}$
as well as $\kappa_{\perp} = \kappafl \sqrt{\kappa_{\parallel} / (2 t_d)}$. Combining the latter two equations in order to eliminate $t_d$ yields
again Eq. (\ref{theimportantformula}). In order to evaluate this further, we consider two sub-cases, namely small and large values of the Kubo number,
respectively. For small Kubo numbers the field line diffusion coefficient is given by the quasi-linear limit
\be
\kappafl = L_{\parallel} \frac{\delta B_x^2}{B_0^2}.
\label{qltkappafl}
\ee
With this form Eq. (\ref{theimportantformula}) becomes
\be
\kappa_{\perp} \approx \left( \frac{L_{\parallel}}{\ell_{\perp}} \right)^2 \frac{\delta B_x^4}{B_0^4} \kappa_{\parallel}
\propto \frac{\ell_{\parallel}^2}{\ell_{\perp}^2} \frac{\delta B_x^4}{B_0^4} \kappa_{\parallel}
\label{smallKuboCLRR}
\ee
in agreement with the scaling obtained from diffusive UNLT theory in Shalchi (2015). Furthermore, we find
\be
L_K \propto \frac{\ell_{\perp}^2}{\ell_{\parallel}} \frac{B_0^2}{\delta B_x^2}.
\label{LKforsmallK}
\ee
For large Kubo numbers, on the other hand, we have
\be
\kappafl = L_{U} \frac{\delta B_x}{B_0}
\label{bohmkappafl}
\ee
with the ultra-scale $L_U$. Eq. (\ref{bohmkappafl}) is either called the non-linear or Bohmian limit and is similar compared to the field line diffusion
coefficient obtained by Kadomtsev \& Pogutse (1979). Therewith, Eq. (\ref{theimportantformula}) becomes
\be
\kappa_{\perp} \approx \left( \frac{L_{U}}{\ell_{\perp}} \right)^2 \frac{\delta B_x^2}{B_0^2} \kappa_{\parallel}
\label{CLRRforlargekubo}
\ee
and the Kolmogorov scale is $L_K = (\ell_{\perp}^2 B_0)/(L_U \delta B_x)$.

\subsection{The Fluid Limit}
Let us now assume that parallel transport is diffusive but magnetic field lines are still ballistic when the particles start to satisfy condition (\ref{thecond}).
Then we can derive $\langle (\Delta x)^2 \rangle =  \langle (\Delta z)^2 \rangle \delta B_x^2 / B_0^2 = 2 \kappa_{\parallel} t \delta B_x^2 / B_0^2$
and, thus,
\be
\kappa_{\perp} = \frac{\delta B_x^2}{B_0^2} \kappa_{\parallel}
\label{fluidlimit}
\ee
which Krommes et al. (1983) called \textit{the fluid limit}.

\subsection{The Initial Free-Streaming Regime}
The simplest case is obtained for the early times when parallel and field line transport are ballistic. In this case
\be
\big< \big( \Delta x \big)^2 \big> = \frac{\delta B_x^2}{B_0^2} \big< \big( \Delta z \big)^2 \big> = \frac{v^2}{3} \frac{\delta B_x^2}{B_0^2} t^2
\label{double1}
\ee
so that
\be
d_{\perp} (t) = \frac{v^2}{3} \frac{\delta B_x^2}{B_0^2} t
\label{double2}
\ee
corresponding to ballistic perpendicular transport. However, this is not a stable regime since we only find this type of transport
before condition (\ref{thecond}) is met.

\subsection{Double-ballistic Diffusion}
We now consider a scenario where the transport is still ballistic when the particles start to satisfy condition (\ref{thecond}). Therefore, we use
Eqs. (\ref{double1}) and (\ref{double2}) to derive $2 \ell_{\perp}^2 = v^2 \delta B_x^2 t_d^2 / (3 B_0^2)$ as well as
$\kappa_{\perp} = v^2 \delta B_x^2 t_d / (3 B_0^2)$. Combining the latter two equations leads to
\be
\kappa_{\perp} = \sqrt{\frac{2}{3}} v \ell_{\perp} \frac{\delta B_x}{B_0}.
\ee
A similar result can be derived from Eq. (\ref{Generalsigmap}) by assuming a ballistic perpendicular motion.

\subsection{Time-scale Arguments}
In order to determine which case is valid for which scenario, one needs to explore at which time a certain process takes place.
In the parallel direction particles need to travel a parallel mean free path in order to get diffusive and, thus,
$t_{\parallel} = \lambda_{\parallel}/v = 3 \kappa_{\parallel} / v^2$. In the following we focus on the case of short $\lambda_{\parallel}$.
For small Kubo numbers the field lines become diffusive for $|z| \approx \ell_{\parallel}$ and the corresponding time is
$t_{\scriptscriptstyle F\!L} \approx \ell_{\parallel}^2 / (2 \kappa_{\parallel})$. Then, on the other hand, if we assume that
condition (\ref{thecond}) is satisfied while the field lines are still ballistic, we have
$2 \ell_{\perp}^2 = 2 \kappa_{\parallel} t_{Fluid} \delta B_x^2 / B_0^2$. For $t_{\parallel} < t_{Fluid} < t_{\scriptscriptstyle F\!L}$
the final state is the fluid limit because then we find that parallel transport becomes diffusive first and then
we meet condition (\ref{thecond}). If, on the other hand, $t_{\parallel} < t_{\scriptscriptstyle F\!L} < t_{Fluid}$ the field lines become diffusive
before condition (\ref{thecond}) is met. This means that we find compound sub-diffusion first. At even later time condition (\ref{thecond}) is eventually
met and diffusion is restored. The corresponding diffusion coefficient is then the CLRR limit. Using the formulas for the times discussed above, this means
that we find CLRR diffusion for $\lambda_{\parallel}^2 \ll \ell_{\parallel}^2 \ll \ell_{\parallel} L_K$ where the Kolmogorov length $L_K$ is given
by Eq. (\ref{LKforsmallK}). Thus for $\lambda_{\parallel} \ll \ell_{\parallel}$ we either find the fluid limit or CLRR diffusion.
If additionally $\ell_{\parallel} \gg L_K$ we find the fluid limit but for $\ell_{\parallel} \ll L_K$ we get CLRR diffusion. It follows from
Eq. (\ref{LKforsmallK}) that $L_K / \ell_{\parallel} \approx K^{-2} \gg 1$ meaning that for small Kubo numbers we always find CLRR diffusion.
For large Kubo numbers similar considerations can be made.

\subsection{A Composite Formula}
A problem of the heuristic approach is that the obtained formulas are only valid in asymptotic limits. Since the two most important cases are CLRR and
FLRW limits, we propose for the perpendicular mean free path defined via $\lambda_{\perp} = 3 \kappa_{\perp} / v$, the formula
\be
\frac{\lambda_{\perp}}{\ell_{\perp}}
= \frac{9 \ell_{\perp}}{16 \lambda_{\parallel}} \left[ \sqrt{ 1 + \frac{8 \kappafl \lambda_{\parallel}}{3 \ell_{\perp}^2} } - 1 \right]^2.
\label{composite}
\ee
Eq. (\ref{composite}) was chosen so that for $\lambda_{\parallel} \rightarrow 0$ we obtain Eq. (\ref{theimportantformula}) and for
$\lambda_{\parallel} \rightarrow \infty$ we get Eq. (\ref{theflrwlimit}). Note that Eq. (\ref{composite}) does not contain the fluid
limit given by Eq. (\ref{fluidlimit}) and, thus, it has some limitations.

\subsection{Further Comments}
The results obtained here are sometimes not comparable to previous results. First of all there are cases such as slab or two-dimensional (2D) turbulence.
In the former case condition (\ref{thecond}) is never satisfied leading to compound sub-diffusion as the final state. In the 2D case parallel transport
is not diffusive (see, e.g., Arendt \& Shalchi (2018)) violating the second rule. In some work (see, e.g., Matthaeus et al. (2003) and Shalchi et al. (2004))
a flat spectrum at large scales was used for the 2D modes. For this type of spectrum the ultra-scale is not finite also violating the second rule.
In order to determine the form of $\kappafl$, we have used the Kubo number. However, in some turbulence models (see, e.g., Goldreich \& Sridhar (1995))
there is only one scale and, thus, the Kubo number becomes $K = \delta B_x / B_0$ often called the \textit{Alfv\'enic Mach number}. The arguments presented
above are still valid.

\section{Comparison Between Theory and Simulations}
As a first example we consider two-component turbulence with dominant 2D modes. For a well-behaving spectrum, Shalchi \& Weinhorst (2009)
have derived
\be
L_U = \sqrt{\frac{s-1}{q-1}} \ell_{\perp}
\label{LUformula}
\ee
requiring $q>1$ for the energy range spectral index and $1<s<2$ for the inertial range spectral index. With the parameter $a^2$ included, non-linear theories
provide in the limit of short parallel mean free paths and 2D turbulence (see, e.g., Shalchi et al. (2004) and Zank et al. (2004))
\be
\kappa_{\perp} = a^2 \frac{\delta B_x^2}{B_0^2} \kappa_{\parallel}.
\label{fluidwitha2}
\ee
According to the heuristic approach we expect CLRR diffusion in the considered parameter regime. Comparing Eqs. (\ref{fluidwitha2}) and (\ref{CLRRforlargekubo})
yields $a = L_U / \ell_{\perp}$ and using Eq. (\ref{LUformula}) for the ultra-scale gives us $a^2 = (s-1)/(q-1)$. Previously it was often assumed that $s=5/3$ and $q=3$
(see for instance Arendt \& Shalchi (2018)) leading to $a^2 = 1/3$. Although it was already stated in Matthaeus et al. (2003) that $a^2 = 1/3$ is needed to achieve
agreement between theory and simulations, in the current paper we found the first time an explanation for this value. It has to be noted that this result was
obtained for a specific form of the spectrum. Alternative spectra and the associated turbulence scales have been discussed in Matthaeus et al. (2007). For some of those
spectra one obtains an ultra-scale larger than the bendover scale. In such cases, however, one would expect that the diffusion coefficient is close to the fluid limit
and, thus, $a^2 \approx 1$.

\begin{figure}[t]
\includegraphics[width=0.48\textwidth]{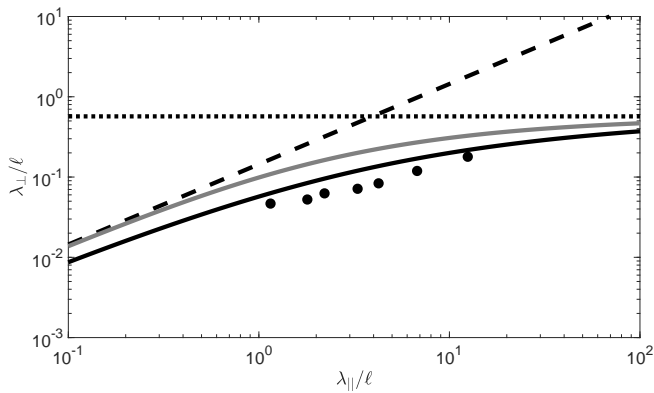}
\caption{Results for a spectral tensor based on the critical balance condition of Goldreich \& Sridhar (1995). For $\delta B / B_0 = 1$ the field line
diffusion coefficient is in this case $\kappa_{FL} = 0.38 \ell$. Shown are the simulations (dots) of Sun \& Jokipii (2011), the result of diffusive
UNLT theory for $a^2 = 1$ (solid line), the CLRR limit (dashed line) as given by Eq. (\ref{theimportantformula}), the FLRW limit (dotted line) as
given by Eq. (\ref{theflrwlimit}), and the composite formula (grey line) as given by Eq. (\ref{composite}).}
\label{GSexample}
\end{figure}
\begin{figure}[t]
\includegraphics[width=0.48\textwidth]{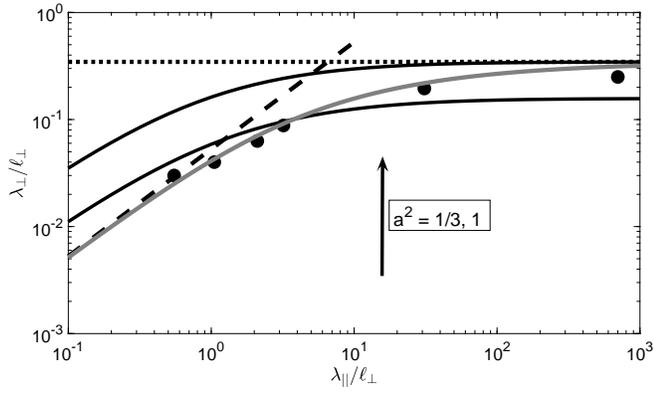}
\caption{Results for a spectral tensor based on the NRMHD model of Ruffolo \& Matthaeus (2013). For $\delta B / B_0 = 1$ the field line diffusion
coefficient is in this case $\kappa_{FL} = 0.23 \ell_{\perp}$. Shown are the simulations (dots) of Shalchi \& Hussein (2014), the results of diffusive
UNLT theory for $a^2 = 1/3$ and $a^2 = 1$ (solid lines), the CLRR limit (dashed line) as given by Eq. (\ref{theimportantformula}), the FLRW limit
(dotted line) as given by Eq. (\ref{theflrwlimit}), and the composite formula (grey line) as given by Eq. (\ref{composite}).}
\label{NRMHDexample}
\end{figure}

Two further examples are shown in Figs. \ref{GSexample} and \ref{NRMHDexample}, respectively. A spectral tensor based on the critical balance condition
of Goldreich \& Sridhar (1995) was used in the simulations of Sun \& Jokipii (2011). Fig. \ref{GSexample} compares diffusive UNLT theory and the heuristic
approach presented in the current paper with those simulations. As shown, the FLRW and CLRR limits have to be understood as asymptotic limits.
Fig. \ref{NRMHDexample} visualizes the comparison for a spectral tensor based on the \textit{noisy reduced MHD (NRMHD) model} of
Ruffolo \& Matthaeus (2013) showing very good agreement.

The heuristic arguments presented in this letter cannot substitute systematic theories due to the lack of accuracy in the general case. The remaining
step is to further improve UNLT theory so that the factor $a^2$ is no longer needed. This should lead to a complete systematic theory for perpendicular
transport in space plasmas.

{}

\end{document}